\documentclass[sigconf,table,xcdraw]{acmart}
\usepackage[ruled,vlined]{algorithm2e}
\usepackage{graphicx}
\usepackage{paralist}
\usepackage{tabularx}
\usepackage{balance}
\usepackage{multirow}
\usepackage{multicol}
\usepackage{pbox}
\usepackage{mathtools}
\usepackage[TABBOTCAP]{subfigure}
\usepackage{algorithmic}
\usepackage{paralist}
\usepackage{tabularx}
\usepackage{url}
\usepackage{balance}
\usepackage{multirow}
\usepackage{multicol}
\usepackage{amsmath}
\usepackage{setspace}
\usepackage{verbatim}
\usepackage{float}
\usepackage[normalem]{ulem}
\usepackage{xspace}
\usepackage{amssymb}
\usepackage{ifthen}
\usepackage{pifont}
\usepackage{tikz}
\newcommand*\circled[1]{\tikz[baseline=(char.base)]{
            \node[shape=circle,draw,inner sep=0.5pt] (char) {#1};}}

\newcommand{\GVT}{{\sc Gvt}\xspace}
\newcommand{\GVTs}{{\sc Gvt's}\xspace}
\newcommand{\dv}{{\textit{DV}}\xspace}
\newcommand{\dvs}{{\textit{DVs}}\xspace}
\newcommand{\gc}{{\textit{GC}}\xspace}
\newcommand{\gcs}{{\textit{GCs}}\xspace}

\include{macro}

\begin{document}

\title{Automated Reporting of GUI Design Violations for Mobile Apps}

\author{Kevin Moran, Boyang Li, Carlos Bernal-C\'ardenas, Dan Jelf, and Denys Poshyvanyk}
\affiliation{%
  \institution{College of William \& Mary \\ Department of Computer Science}
  \streetaddress{P.O. Box 8795}
  \city{Williamsburg} 
  \state{VA, USA} 
  \postcode{23185}
}
\email{{kpmoran,boyang,cebernal,dkjelf,denys}@cs.wm.edu}

\renewcommand{\shortauthors}{K. Moran, B. Li, C. Bernal-C\'ardenas, D. Jelf, and D. Poshyvanyk}

\begin{abstract}
	The inception of a mobile app often takes form of a mock-up of the Graphical User Interface (GUI), represented as a static image delineating the proper layout and style of GUI widgets that satisfy requirements.  Following this initial mock-up, the design artifacts are then handed off to developers whose goal is to accurately implement these GUIs and the desired functionality in code.  
Given the sizable abstraction gap between mock-ups and code, developers often introduce mistakes related to the GUI that can negatively impact an app's success in highly competitive marketplaces.  Moreover, such mistakes are common in the evolutionary context of rapidly changing apps. This leads to the time-consuming and laborious task of design teams verifying that each screen of an app was implemented according to intended design specifications.

	This paper introduces a novel, automated approach for verifying whether the GUI of a mobile app was implemented according to its intended design. Our approach resolves GUI-related information from both implemented apps and mock-ups and uses computer vision techniques to identify common errors in the implementations of mobile GUIs.  We implemented this approach for Android in a tool called \GVT and carried out both a controlled empirical evaluation with open-source apps as well as an industrial evaluation with designers and developers from Huawei. The results show that \GVT solves an important, difficult, and highly practical problem with remarkable efficiency and accuracy and is both useful and scalable from the point of view of industrial designers and developers. The tool is currently used by over one-thousand industrial designers \& developers at Huawei to improve the quality of their mobile apps.
\vspace{-0.1cm}
\end{abstract}


\copyrightyear{2018}
\acmYear{2018} 
\setcopyright{acmlicensed}
\acmConference[ICSE '18]{ICSE '18: 40th International Conference on Software Engineering }{May 27-June 3, 2018}{Gothenburg, Sweden}
\acmBooktitle{ICSE '18: ICSE '18: 40th International Conference on Software Engineering , May 27-June 3, 2018, Gothenburg, Sweden}
\acmPrice{15.00}
\acmDOI{10.1145/3180155.3180246}
\acmISBN{978-1-4503-5638-1/18/05}

\maketitle



\vspace{-0.1cm}
\section{Introduction}
\label{sec:intro}

	Intuitive, elegant graphical user interfaces (GUIs) embodying effective user experience (UX) and user interface (UI) design principles are essential to the success of mobile apps.  In fact, one may argue that these design principles are largely responsible for launching the modern mobile platforms that have become so popular today.  Apple Inc's launch of the iPhone in 2007 revolutionized the mobile handset industry (heavily influencing derivative platforms including Android) and largely centered on an elegant, well-thought out UX experience, putting multitouch gestures and a natural GUI at the forefront of the platform experience. A decade later, the most successful mobile apps on today's highly competitive app stores (\eg Google Play\cite{google-play} and Apple's App Store\cite{apple-app-store}) are those that embrace this focus on ease of use, and blend intuitive user experiences with beautiful interfaces.  In fact, given the high number of apps in today's marketplaces that perform remarkably similar functions \cite{app-abandonment}, the design and user experience of an app are often differentiating factors, leading to either success or failure \cite{design-importance}. 

	Given the importance of a proper user interface and user experience for mobile apps, development usually begins with UI/UX design experts creating highly detailed mock-ups of app screens using one of several different prototyping techniques \cite{Silva:AGILE11,Kuusinen:APSEC13}.  The most popular of these techniques and the focus of this paper, is referred to as \textit{mock-up driven development} where a designer (or group of designers) creates pixel perfect representations of app UIs using software such as Sketch\cite{sketch} or PhotoShop\cite{photoshop}. Once the design artifacts (or \textit{mock-ups}) are completed, they are handed off to development teams who are responsible for implementing the designs in code for a target platform.  In order for the design envisioned by the UI/UX experts (who carry domain knowledge that front-end developers may lack) to be properly transferred to users, an accurate translation of the mock-up to code is \textit{essential}.

	Yet, implementing an intuitive and visually appealing UI in code is well-known to be a challenging undertaking \cite{Tucker:CSH04,Myers:CHD94,Nguyen:ASE15}.  As such, many mobile development platforms such as Apple's Xcode IDE and Android Studio include powerful built-in GUI editors.  Despite the ease of use such technologies are intended to facilitate, a controlled study has illustrated that such interface builders can be difficult to operate, with users prone to introducing bugs \cite{Zeidler:INTERACT13}.  Because apps under development are prone to errors in their GUIs, this typically results in an iterative workflow where UI/UX teams will frequently \textit{manually audit} app implementations during the development cycle and report any violations to the engineering team who then aims to fix them.  This incredibly time consuming back-and-forth process is further complicated by several underlying challenges specific to mobile app development including: (i) continuous pressure for frequent releases \cite{Hu:ESYS14,Jones:2014}, (ii) the need to address user reviews quickly to improve app quality \cite{Palomba:ICSME15,Palomba:ICSE17,Ciurumelea:SANER17,DiSorbo:FSE16}, (iii) frequent platform updates and API instability \cite{Bavota:TSE15,Linares:FSE13,Linares:ICPC14,ICSM13:KIM} including changes in UI/UX design paradigms inducing the need for GUI re-designs (\eg material design), and (iv) the need for custom components and layouts to support complex design mock-ups.  Thus, there is a practical need for  effective automated support to improve the process of detecting and reporting design violations and providing developers with more accurate and actionable information.	

	The difficulty that developers experience in creating effective GUIs stems from the need to manually bridge a staggering abstraction gap that involves reasoning concise and accurate UI code from pixel-based graphical representations of GUIs. The GUI errors that are introduced when attempting to bridge this gap are known in literature as \textit{presentation failures}. Presentation failures have been defined in the context of web applications in previous work as \textit{``a discrepancy between the actual appearance of a webpage [or mobile app screen] and its intended appearance"} \cite{Mahajan:ICST16}.   We take previous innovative work that aims to detect presentation errors in web applications \cite{Mahajan:ICST15,Mahajan:ICST16,RoyChoudhary:ICSE13,Choudhary:ICST12} as motivation to design equally effective approaches in the domain of mobile apps.  Presentation failures are typically comprised of several \textit{visual symptoms} or specific mismatches between visual facets of the intended GUI design and the implementation of those GUI-components \cite{Mahajan:ICST16} in an app.   These visual symptoms can vary in type and frequency depending on the domain (\eg web vs. mobile), and in the context of mock-up driven development, we define them as \textit{design violations}. 

	In this paper, we present an approach, called \GVT (\textbf{G}ui \textbf{V}erification sys\textbf{T}em), developed in close collaboration with Huawei. Our approach is capable of automated, precise reporting of the design violations that induce presentation failures between an app mock-up and its implementation.  Our technique decodes the hierarchal structure present in both mockups and dynamic representations of app GUIs, effectively matching the corresponding components. \GVT then uses a combination of computer vision techniques to accurately detect design violations. Finally, \GVT constructs a report containing screenshots, links to static code information (if code is provided ), and precise descriptions of design violations.  \textit{GVT was developed to be practical and scalable, was built in close collaboration with the UI/UX teams at Huawei, and is currently in use by over one-thousand designers and engineers at the company.}

	To evaluate the performance and usefulness of \GVT we conducted three complementary studies.  First, we empirically validated \GVTs \textit{performance}  by measuring the precision and recall of detecting synthetically injected design violations in popular open source apps.  Second, we conducted a user study to measure the \textit{usefulness} of our tool, comparing \GVTs ability to detect and report design violations to the ability of developers, while also measuring the perceived utility of \GVT reports.  Finally, to measure the \textit{applicability} of our approach in an industrial context, we present the results of an industrial case study including: (i) findings from a survey sent to industrial developers and designers who use \GVT in their development workflow and (ii) semi-structured interviews with both design and development team managers about the impact of the tool.  Our findings from this wide-ranging evaluation include the following key points: (i) In our study using synthetic violations \GVT is able to detect design violations with an overall precision of 98\% and recall of 96\%; (ii) \GVT is able to outperform developers with Android development experience in identifying design violations while taking less time; (iii) developers generally found \GVTs reports useful for quickly identifying different types of design violations; and (iv) \GVT had a meaningful impact on the design and development of mobile apps for our industrial partner, contributing to increased UI/UX quality.

Our paper contributions can be summarized as follows:

\begin{itemize}
	\item{We formalize the concepts of \textit{presentation failures} and \textit{design violations} for mock-up driven development in the domain of mobile apps, and empirically derive common types of design violations in a study on an industrial dataset;}

	\item{We present a novel approach for detecting and reporting these violations embodied in a tool called \GVT that uses hierarchal representations of an app's GUI and computer vision techniques to detect and accurately report \textit{design violations;}}
	
	\item{We conduct a wide-ranging evaluation of the \GVT studying its \textit{performance}, \textit{usefulness}, and industrial \textit{applicability};}
	
	\item{We include an online appendix \cite{appendix} with examples of reports generated by \GVT and our evaluation dataset. Additionally, we make the \GVT tool and code available upon request.}
\end{itemize}


\vspace{-0.3cm}
\section{Problem Statement \& Origin}
\label{sec:background}

	In this section we formalize the problem of detecting \textit{design violations} in GUIs of mobile apps and discuss the origin of the problem rooted in industrial mobile app design \& development.

\vspace{-0.3cm}
\subsection{Problem Statement}
\label{subsec:prob-statement}

	At a high level, our goal is to develop an automated approach capable of detecting, classifying, and accurately describing \textit{design violations} that exist for a single screen of a mobile app to help developers resolve \textit{presentation failures} more effectively. In this section we formalize this scenario in order to allow for an accurate description and scope of our proposed approach.  While this section focuses on concepts, Sec. \ref{sec:approach} focuses on the implementation details.

\vspace{-0.2cm}
\subsubsection{GUI-Components \& Screens}
\label{subsubsec:guis-screens}

	There are two main logical constructs that define the concept of the GUI of an app: \textit{GUI-components}  (or GUI-widgets) and \textit{Screens}.  A \textit{GUI-component} is a discrete object with a set of attributes (such as size and location among others) associated with a particular \textit{Screen} of an app.  A \textit{Screen} is an invisible canvas of size corresponding to the physical screen dimensions of a mobile device. We define two types of screens, those created by designers using professional-grade tools like Sketch, and those collected from implemented apps at runtime.  Each of these two types of Screens has an associated set of GUI-components (or \textit{components}). Each set of components associated with a screen is structured as a cumulative hierarchy comprising a tree structure, starting with a single root node, where the spatial layout of parent always encompasses contained child components.

\noindent \textit{\textbf{Definition 1: GUI-Component (GC) -}} A discrete object $GC$ with a corresponding set of attributes $a$ which can be represented as a four-tuple in the form \textit{(<x-position,y-position>, <height,width>, <text>, <image>)}.  Here the first four elements of the tuple describe the location of the top left point for the bounding box of the component, and the height and width attributes describe the size of the bounding box. The text attribute corresponds to text displayed by the component and the image attribute represents an image of the component with bounds adhering to the first two attributes.

\noindent \textit{\textbf{Definition 2: Screen (S) -}} A canvas $S$ with a predefined height and width corresponding to the physical display dimensions of a smartphone or tablet.  Each Screen contains a cumulative hierarchy of components, which can be represented as a nested set such that:
\begin{equation}
\label{eq:screen-set}
S = \{GC_1\{GC_2\{GC_i\}, GC_3\}\}
\end{equation}
\noindent where each $GC$ has a unique attribute tuple and the nested set can be ordered in either depth-first (Exp. \ref{eq:screen-set}) or in a breadth-first manner. We are concerned with two specific types of screens: screens representing mock-ups of mobile apps $S^m$ and screens representing real implementations of these apps, or $S^r$.

\begin{figure}
\centering
\vspace{-0.2cm}
\includegraphics[width=\columnwidth]{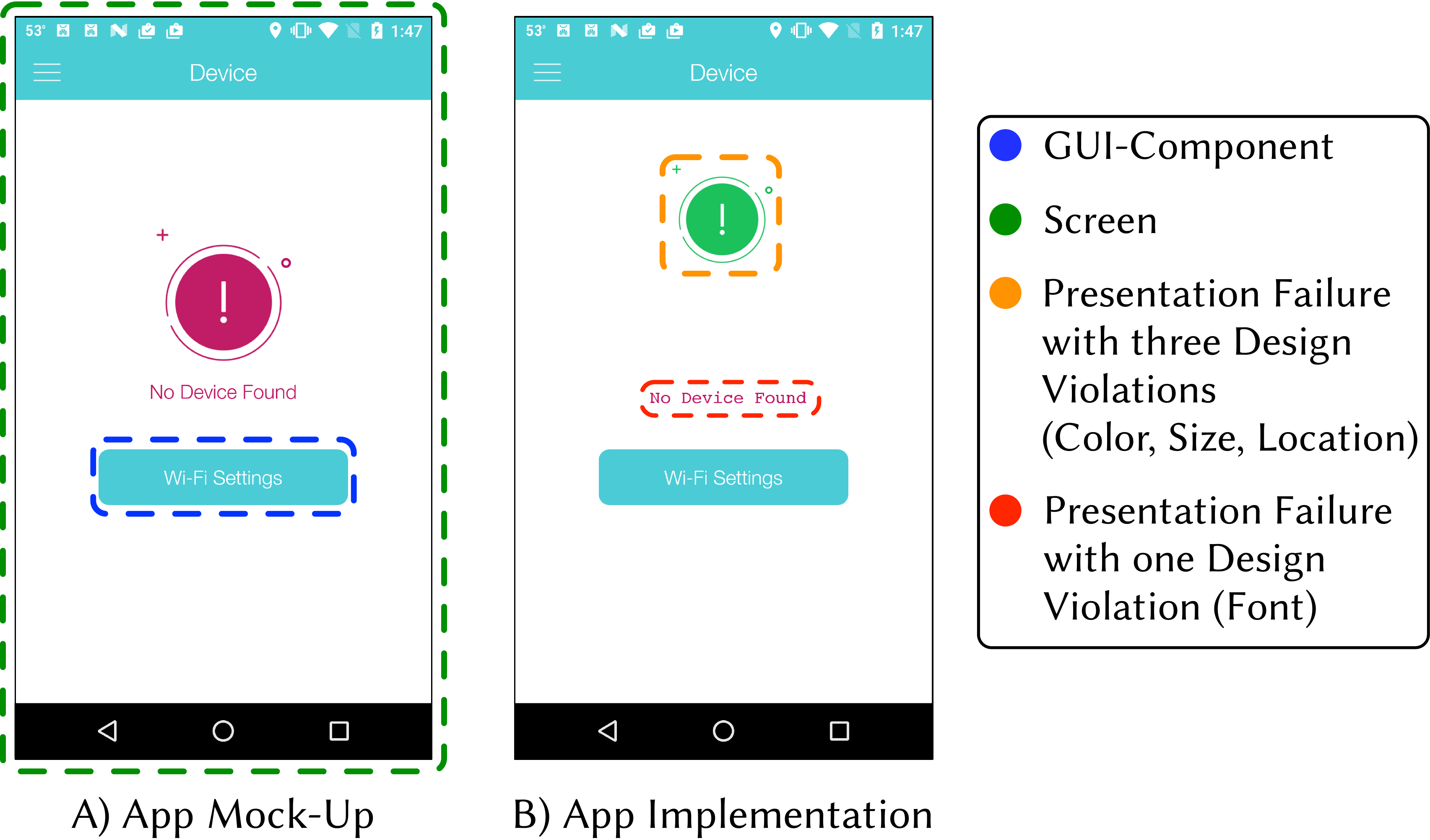}
\vspace{-0.0cm}
\centering
\vspace{-0.6cm}
\caption{Examples of Formal Definitions}
\label{fig:def-example}
\end{figure}

\subsubsection{Design Violations \& Presentation Failures}
\label{subsubsec:violations-failures}

	As described earlier, \textit{design violations} correspond to visual symptoms of \textit{presentation failures}, or differences between the intended design and implementation of a mobile app screen. Presentation failures can be made up of one or more design violations of different types.

\noindent \textit{\textbf{Definition 3: Design Violation (DV) - }} As shown in Exp. \ref{eq:design-violation}, a mismatch between the attribute tuples of two corresponding leaf-level (\ie having no direct children) GUI-components $GC_{i}^m$ and $GC_{j}^r$ of two screens $S^m$ and $S^r$ imply a design violation \textit{DV} associated with those components. In this definition leaf nodes \textit{correspond} to one another if their location and size on a screen (\ie \textit{<x-position,y-position>, <height,width>}) match within a given threshold.  \textit{Equality} between leaf nodes is measured as a tighter matching threshold across all attributes. As we illustrate in the next section, inequalities between different attributes in the associated tuples of the \gcs lead to different types of design violations.

\begin{equation}
\begin{aligned}
\label{eq:design-violation}
\ (GC_{i}^m \approx GC_{j}^r) \wedge (GC_{i}^m \neq GC_{j}^r)\\ \implies DV \in \{GC_{i}^m,GC_{j}^r\}
\end{aligned}
\end{equation}

\noindent \textit{\textbf{Definition 4: Presentation Failure \textit{(PF)} -}} A set of one or more \dvs attributed to a set of corresponding GUI-components between two screens $S_m$ and $S_r$, as shown in Exp. 3. For instance, as shown in Fig. \ref{fig:def-example}, a single set of corresponding components may have differences in both the \textit{<x,y>} and \textit{<height,width>} attributes leading to two constituent design violations that induce a single presentation failure \textit{PF}. Thus, each presentation failure between two Screens $S$ corresponds to at least one mismatch between the attribute vectors of two corresponding leaf node GUI-components $GC_{im}$ and $GC_{ir}$.

\begin{equation}
\begin{aligned}
\label{eq:presentation-failure}
\textrm{if} \quad \{DV_1, DV_2,... DV_i\} \in \{GC_{i}^m,GC_{j}^r\}\\
\textrm{then} \quad PF \in \{S^m, S^r\}
\end{aligned}
\end{equation}

\subsubsection{Problem Statement}
\label{subsubsec:guis-screens}

Given these definitions, the problem being solved in this paper is the following:  Given two screens $S^m$ and $S^r$ corresponding to the mock-up and implementation screens of a mobile application, we aim to detect and describe the set of presentation failures $\{PF_1,PF_2,... PF_i \} \in \{S^m, S^r\}$.  Thus, we aim to report all design violations on corresponding \gc pairs: 

\vspace{-0.25cm}
\begin{equation}
\begin{aligned}
\{DV_1,DV_2,... DV_k \} \in \\
\{\{GC_{i_1}^m,GC_{j_1}^r\}, \{GC_{i_2}^m,GC_{j_2}^r\},... \{GC_{i_x}^m,GC_{j_y}^r\}\}
\end{aligned}
\end{equation}

\subsection{Industrial Problem Origins}
\label{subsec:problem-origin}
 
	A typical industrial mobile development process includes the following steps (as confirmed by our collaborators at Huawei): (i) First a team of designers creates highly detailed mockups of an app's screens using the Sketch \cite{sketch} (or similar) prototyping software. These mock-ups are typically ``pixel-perfect" representations of the app for a given screen dimension; (ii) The mock-ups are then handed off to developers in the form of exported images with designer added annotations stipulating spatial information and constraints. Developers use this information to implement representations of the GUIs for Android using a combination of Java and \texttt{xml}; (iii) Next, after the initial version of the app has been implemented, compiled Android Application Package(s) (\ie \texttt{\small apks}) are sent back to the designers who then install these apps on target devices, generate screenshots for the screens in question, and manually search for discrepancies compared to the original mock-ups; (iv) Once the set of violations are identified, these are communicated back to the developers via textual descriptions and annotated screenshots at the cost of significant manual effort from the design teams. Developers must then identify and resolve the \dvs using this information.  The process is often repeated in several iterations causing substantial delays in the development process.

	The goal of our work is to drastically improve this iterative process by: (i) automating the identification of \dvs on the screens of mobile apps - saving both the design and development teams time and effort, and (ii) providing highly accurate information to the developers regarding these \dvs in the form of detailed reports - in order to reduce their effort in resolving the problem.

\vspace{-0.3cm}
\section{Design Violations in Practice}
\label{sec:motivating-study}

	In order to gain a better understanding of the types of \dvs that occur in mobile apps in practice, we conducted a study using a dataset from Huawei. While there do exist a small collection of taxonomies related to visual GUI defects \cite{Lelli:ICST15,ISSA:WSE12} and faults in mobile apps \cite{Hall:EUROMICRO:14,Linares-Vasquez:FSE17}, we chose to conduct a contextualized study with our industrial partner for the following reasons: (i) existing taxonomies for visual GUI defects were not detailed enough, containing only general faults (\eg ``incorrect appearance''), (ii) existing fault taxonomies for mobile apps either did not contain visual GUI faults or were not complete, and (iii) we wanted to derive a contextualized \dv taxonomy for apps developed at Huawei. The findings from this study underscore the existence and importance of the problem that our approach aims to solve in this context.  Due to an NDA, we are not able to share the dataset or highlight specific examples, in order to avoid revealing information about future products at Huawei.  However, we present aggregate results in this section.

\vspace{-0.2cm}
\subsection{Study Setting \& Methodology}
\label{subsec:study-methods}

	The \textit{goal} of this study is to derive a taxonomy of the different types of \dvs and examine the distribution of these types induced during the mobile app development process.  The \textit{context} of this study is comprised of a set of 71 representative mobile app mock-up and implementation screen pairs from more than 12 different internal apps, annotated by design teams from our industrial partner to highlight specific instances of \textit{resolved} \dvs.  This set of screen pairs was specifically selected by the industrial design team to be representative both in terms of diversity and distribution of violations that typically occur during the development process.

	In order to develop a taxonomy and distribution of the violations present in this dataset, we implement an open coding methodology consistent with constructivist grounded theory \cite{Charmaz:groundedtheory}. Following the advice of recent work within the SE community \cite{Stol:ICSE16}, we stipulate our specific implementation of this type of grounded theory while discussing our deviations from the methods in the literature. We derived our implementation from the material discussed in \cite{Charmaz:groundedtheory} involving the following steps: (i) establishing a research problem and questions, (ii) data-collection and initial coding, and (iii) focused coding.  We excluded other steps described in \cite{Charmaz:groundedtheory}, such as memoing because we were building a taxonomy of labels, and seeking new specific data due to our NDA limiting the data that could be shared. 
	The study addressed the following research question: \textit{What are the different types and distributions of GUI design violations that occur during industrial mobile app development processes?}
	
\begin{figure}[t]
\centering
\vspace{-0.3cm}
\includegraphics[width=0.85\columnwidth]{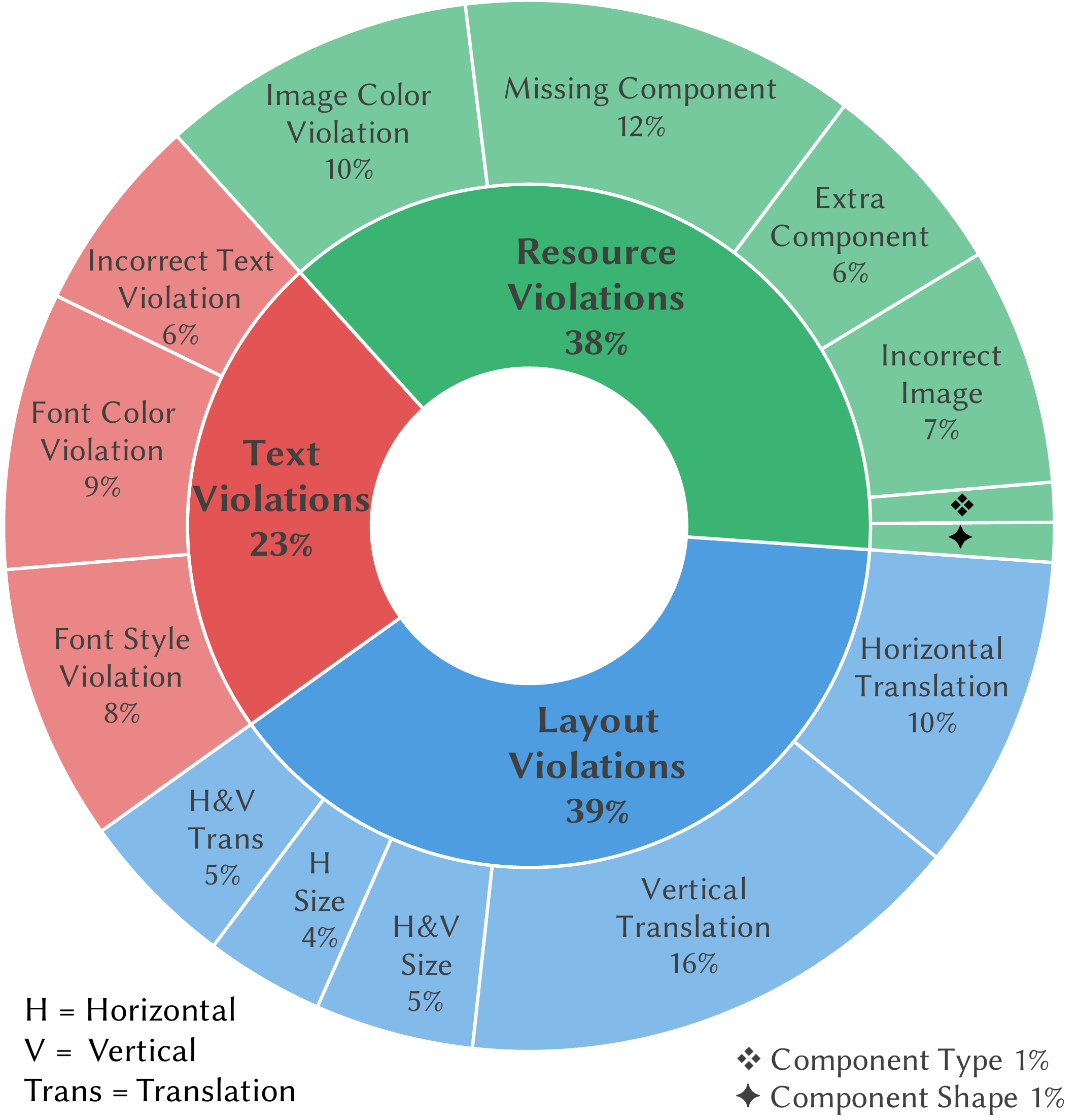}
\vspace{-0.0cm}
\centering
\vspace{-0.35cm}
\caption{Distribution of Different Types of Industrial \dvs}
\label{fig:violation-distribution}
\end{figure}

	During the initial coding process, three of the authors were sent the full set of 71 screen pairs and were asked to code four pieces of information for each example: (i) a general category for the violation, (ii) a specific description of the violation, (iii) the severity of the violation (if applicable), and (iv) the Android \gc types affected (\eg button).  Finally, we performed a second round of coding that combined the concepts of focused and axial coding as described in \cite{Charmaz:groundedtheory}. During this round two of the authors merged the responses from all three types of coding information where at least two of the three coders agreed. During this phase similar coding labels were merged (\eg ``layout violation" vs. ``spatial violation"), conflicts were resolved, two screen pairs were discarded due to ambiguity, and cohesive categories and subcategories were formed.  The author agreement for each of the four types of tags is as follows: (i) general violation category (100\%), (ii) specific violation description (96\%), (iii) violation severity (100\%), and (iv) affected \gc types (84.5\%).

\vspace{-0.3cm}
\subsection{Grounded Theory Study Results}
\label{subsec:study-results}
	
	Our study revealed three major categories of design violations, each with several specific subtypes.  We forgo detailed descriptions and examples of violations due to space limitations, but provide examples in our online appendix \cite{appendix}. 
The derived categories and subcategories of \dvs, and their distributions, are illustrated in Fig.~\ref{fig:violation-distribution}. Overall 82 \dvs were identified across the 71 unique screen pairs considered in our study. The most prevalent category of \dvs in our taxonomy are \textit{Layout Violations} ($\approx 40\%$), which concern either a translation of a component in the \textit{x} or \textit{y} direction or a change in the component size, with translations being more common. The second most prevalent category ($\approx 36\%$) consists of \textit{Resource Violations}, which concern missing components, extra components, color differences, and image differences. Finally, about one-quarter ($\approx 24\%$) of these violations are \textit{Text Violations}, which concern differences in components that display text. We observed that violations typically only surfaced for ``leaf-level" components in the GUI hierarchy. That is, violations typically only affected atomic components \& not containers or backgrounds.  Only 5/82 of examined violations ($\approx 6\%$) affected backgrounds or containers. Even in these few cases, the violations also affected ``leaf-level" components.

	The different types of violations correspond to different inequalities between the attribute tuples of corresponding GUI-components defined in Sec. \ref{sec:background}. This taxonomy shows that designers are charged with identifying several different types of design violations, a daunting task, particularly for hundreds of screens across several apps.

\vspace{-0.2cm}
\section{The \GVT Approach}
\label{sec:approach}
\vspace{-0.1cm}

\begin{figure*}[tb]
\centering
\vspace{-0.5cm}
\includegraphics[width=\linewidth]{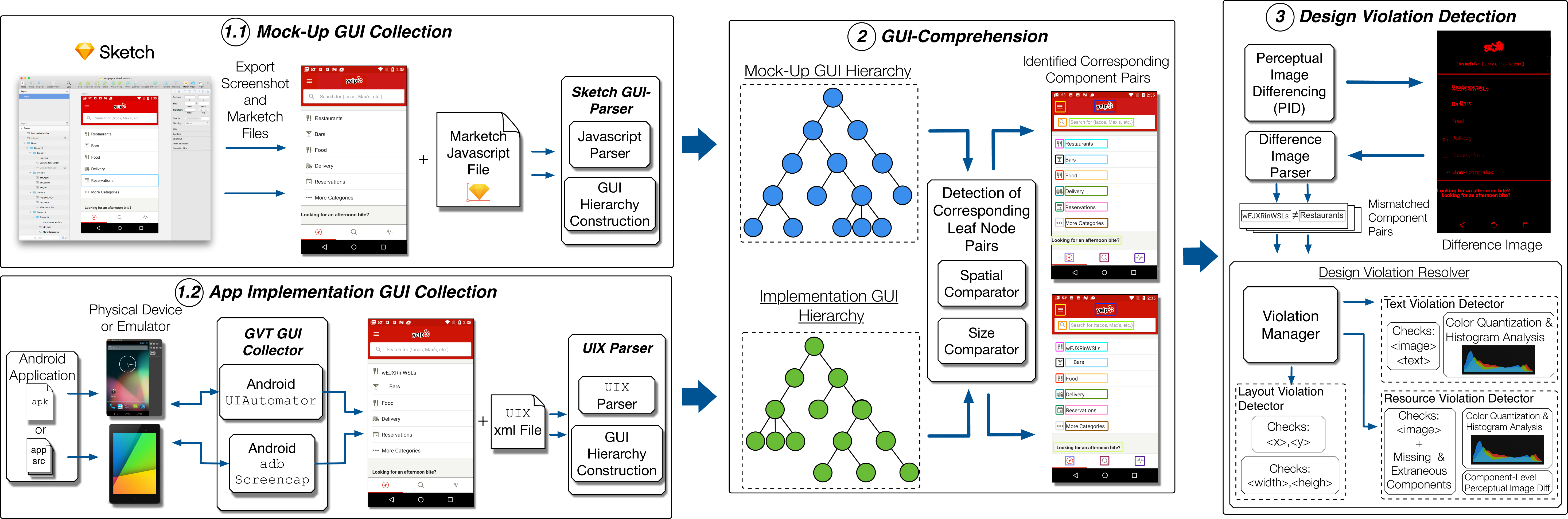}
\vspace{-0.8cm}
\caption{Overview of GVT Workflow}
\label{fig:design}
\vspace{-0.5cm}
\end{figure*}

\subsection{Approach Overview}
\label{subsec:approach-overview}

	The workflow of \GVT (Fig. \ref{fig:design}) proceeds in three stages: First in the \textit{GUI-Collection Stage}, GUI-related information from both mock-ups and running apps is collected; Next, in the \textit{GUI-Comprehension Stage} leaf-level \gcs are parsed from the trees and a KNN-based algorithm is used to match corresponding \gcs using spatial information; Finally, in the \textit{Design Violation Detection Stage} \dvs are detected using a combination of methods that leverage spatial \gc information and computer vision techniques.

\vspace{-0.3cm}
\subsection{Stage 1: GUI Collection}
\label{subsec:gui-collection}

\subsubsection{Mock-Up GUI Collection}
\label{subsubsec:mockup-gui-collection}

	Software UI/UX design professionals typically use professional-grade image editing software (such as Photoshop\cite{photoshop} or  Sketch\cite{sketch}) to create their mock-ups.  Designers employed by our industrial partner utilize the Sketch design software.  Sketch is popular among mobile UI/UX designers due to its simple but powerful features, ease of use, and large library of extensions \cite{sketch-ext}. When using these tools designers often construct graphical representations of smartphone applications by placing \textit{objects} representing \gcs (which we refer to as \textit{mock-up} \gcs) on a \textit{canvas} (representing a Screen \textit{S}) that matches the typical display size of a target device.  In order to capture information encoded in these mock-ups we decided to leverage an export format that was already in use by our industrial partner, an open-source Sketch extension called Marketch \cite{marketch} that exports mock-ups as an \texttt{\small html} page including a screenshot and \texttt{\small JavaScript} file.

	Thus, as input from the mock-up, \GVT receives a screenshot (to be used later in the \textit{Design Violation Detection Phase}) and a directory containing the Marketch information. The \texttt{\small JavaScript} file contains several pieces of information for each mock-up \gc including, (i) the location of the mock-up \gc on the canvas, (ii) size of the bounding box, and (iii) the text/font displayed by the mock-up \gc (if any).  As shown in Figure \ref{fig:design}-\circled{1.1}, we built a parser to read this information. \textit{However, it should be noted that our approach is not tightly coupled to Sketch or Marketch files.}\footnote{Similar information regarding mock-up \gcs can be parsed from the \texttt{\footnotesize html} or Scalable Vector Graphics (\texttt{\footnotesize .svg}) format exported by other tools such as Photoshop\cite{photoshop}.}  After the Marketch files have been parsed, \GVT examines the extracted spatial information to build a \gc hierarchy. The result can be logically represented as a rooted tree where leaf nodes contain the atomic UI-elements with which a typical user might interact. Non-leaf node components typically represent containers, that form logical groupings of leaf node components and other containers. In certain cases, our approximation of using mock-up \gcs to represent implementation \gcs may not hold. For instance, an icon which should be represented as a single \gc may consist of several mock-up \gcs representing parts of the icon.  \GVT handles such cases in the \textit{GUI-Comprehension} stage.

\vspace{-0.15cm}
\subsubsection{Dynamic App GUI-Collection}
\label{subsubsec:impl-gui-collection}

	In order to compare the  the mock-up of an app to its implementation \GVT must extract GUI-related meta-data from a running Android app. \GVT is able to use Android's  \texttt{\small uiautomator} framework~\cite{uiautomator} intended for UI testing to capture \texttt{\small xml} files and screenshots for a target screen of an app running on a physical device or emulator. Each \texttt{\small uiautomator} file contains information related to the runtime GUI-hierarchy of the target app, including the following attributes utilized by \GVT: (i) The Android component type (\eg \texttt{\small android.widget.ImageButton}), (ii) the location on the screen, (iii) the size of the bounding box, (iv) text displayed, (v) a developer assigned \texttt{\small id}.  The hierarchal structure of components is encoded directly in the \texttt{\small uiautomator} file, and thus we built a parser to extract GUI-hierarchy using this information directly (see Fig. \ref{fig:design}-\circled{1.2}).

\vspace{-0.15cm}
\subsection{Stage 2: GUI Comprehension}
\label{subsec:gui-comprehension}
\vspace{-0.15cm}

	In order for \GVT to find visual discrepancies between components existing in the mock-up and implementation of an app, it must determine which components correspond to one another.  Unfortunately, the GUI-hierarchies parsed from both the Marketch, and \texttt{\small uiautomator} files tend to differ dramatically due to several factors, making tree-based \gc matching difficult. First, since the hierarchy constructed using the Marketch files is generated using information from the Sketch mock-up of app, it is using information derived from designers.  While designers have tremendous expertise in constructing visual representations of apps, they typically do not take the time to construct programmatically-oriented groupings of components. Furthermore, designers are typically not aware of the correct Android component types that should be attributed to different objects in a mock-up.  Second, the \texttt{\small uiautomator} representation of the GUI-hierarchy contains the runtime hierarchal structure of \gcs and correct \gc types.  This tree is typically far more complex, containing several levels of containers grouping \gcs together, which is required for the responsive layouts typical of mobile apps. 

	To overcome this challenge, \GVT instead forms two collections of \textit{leaf-node} components from both the mock-up and implementation GUI-hierarchies (Fig. \ref{fig:design}-\circled{2}), as this information can be easily extracted.  As we reported in Sec. \ref{sec:motivating-study}, the vast majority of \dvs affects leaf-node components.  Once the leaf node components have been extracted from each hierarchy, GVT employs a K-Nearest-Neighbors (KNN) algorithm utilizing a similarity function based on the location and size of the \gcs in order to perform matching.  In this setting, an input leaf-node component from the mock-up would be matched against it closest (\eg K=1) neighbor from the implementation based upon the following similarity function:
\begin{equation}
\label{eq:MatchingFunction}
\gamma = (|x_m - x_r| + |y_m - y_r| + |w_m - w_r| + |h_m - h_r|)
\end{equation}
Where $\gamma$ is a similarity score where smaller values represent closer matches. The $x,y,w$ and $h$ variables correspond to the $x$ \& $y$ location of the top and left-hand borders of the bounding box, and the height and width of the bounding boxes for the mock-up and implementation \gcs respectively.  The result is a list of \gcs that should logically correspond to one another (\textit{corresponding \gcs}).  

	It is possible that there exist instances of missing or extraneous components between the mock-up and implementation.  To identify these cases, our KNN algorithm employs a \textit{GC-Matching Threshold} ($MT$).  If the similarity score of the nearest neighbor match for a given input mock-up \gc exceeds this threshold, it is not matched with any component, and will be reported as a \textit{missing} \gc violation.  If there are unmatched \gcs from the implementation, they are later reported as \textit{extraneous} \gc violations.

	Also, there may be cases where a logical \gc in the implementation is represented as small group of mock-up \gcs.  \GVT is able to handle these cases using the similarity function outlined above.  For each mock-up \gc, \GVT checks whether the neighboring \gcs in the mockup are closer than the closest corresponding \gc in the implementation. If this is the case, they are merged, with the process repeating until a logical GUI-component is represented.

\vspace{-0.3cm}
\subsection{Stage 3: Design Violation Detection}
\label{subsec:violation-detection}

	In the \textit{Design Violation Detection} stage of the \GVT workflow, the approach uses a combination of computer vision techniques and heuristic checking in order to effectively detect the different categories of \dvs derived in our taxonomy presented in Section \ref{sec:motivating-study}.

\vspace{-0.2cm}
\subsubsection{Perceptual Image Differencing}
\label{subsubsec:percep-diff}

     In order to determine corresponding \gcs with visual discrepancies \GVT uses a technique called Perceptual Image Differencing (PID) \cite{Yee:TG01} that operates upon the mock-up and implementation screenshots.  PID utilizes a model of the human visual system to compare two images and detect visual differences, and has been used to successfully identify visual discrepancies in web applications in previous work \cite{Mahajan:ICST15,Mahajan:ICST16}. We use this algorithm in conjunction with the \gc information derived in the previous steps of \GVT to achieve accurate violation detection. For a full description of the algorithm, we refer readers to \cite{Yee:TG01}.  The PID algorithm uses several adjustable parameters including: $F$ which corresponds to the visual field of view in degrees, $L$ which indicates the luminance or brightness of the image, and $C$ which adjusts sensitivity to color differences. The values used in our implementation are stipulated in Section \ref{subsec:impl-collab}.

	The output of the PID algorithm is a single \textit{difference image} (Fig. \ref{fig:design}-\circled{3}) containing \textit{difference pixels}, which are pixels considered to be perceptually different between the two images.  After processing the difference image generated by PID, \GVT extracts the implementation bounding box for each corresponding pair of \gcs, and overlays the box on top of the generated difference image. It then calculates the number of difference pixels contained within the bounding box where higher numbers of difference pixels indicate potential visual discrepancies. Thus, \GVT collects all ``suspicious" \gc pairs with a \% of difference pixels higher than a \textit{Difference Threshold} $DT$. This set of suspicious components is then passed to the \textit{Violation Manager} (Fig. \ref{fig:design}-\circled{3}) so that specific instances of \dvs can be detected.

\vspace{-0.2cm}
\subsubsection{Detecting Layout Violations}
\label{subsubsec:layoout-violations}

	The first general category of \dvs that \GVT detects are \textit{Layout Violations}. According the taxonomy derived in Sec. \ref{sec:motivating-study} there are six specific layout \dv categories that relate to two component properties: (i) screen location (\ie \textit{<x,y>} position) and (ii) size (\ie \textit{<h,w>} of the \gc bounding box). \GVT first checks for the three types of translation \dvs utilizing a heuristic that measures the distance from the top and left-hand edges of matched components. If the difference between the components in either the $x$ or $y$ dimension is greater than a \textit{Layout Threshold} ($LT$), then these components are reported as a \textit{Layout} \dv. Using the $LT$ avoids trivial location discrepancies within design tolerances being reported as violations, and can be set by a designer or developer using the tool.  When detecting the three types of size \dvs in the derived design violation taxonomy, \GVT utilizes a heuristic that compares the width and height of the bounding boxes of corresponding components.  If the width or height of the bounding boxes differ by more than the $LT$, then a layout violation is reported.

\vspace{-0.2cm}
\subsubsection{Detecting Text Violations}
\label{subsubsec:text-violations}

	The next general type of \dv that \GVT detects are \textit{Text Violations}, of which there are three specific types: (i) Font Color, (ii) Font Style, and (iii) Incorrect Text Content.   These detection strategies are only applied to pairs of text-based components as determined by \texttt{\small uiautomator} information. To detect font color violations, \GVT extracts cropped images for each pair of suspicious text components by cropping the mock-up and implementation screenshots according to the component's respective bounding boxes.  Next,  \textit{Color Quantization (CQ)} is applied to accumulate instances of all unique RGB values expressed in the component-specific images.  This quantization information is then used to construct a \textit{Color Histogram (CH)} (Fig. \ref{fig:design}-\circled{3}).  \GVT computes the normalized Euclidean distance between the extracted Color Histograms for the corresponding \gc pairs, and if the Histograms do not match within a \textit{Color Threshold (CT)} then a \textit{Font-Color} \dv is reported and the top-3 colors (i.e, centroids) from each CH are recorded in the \GVT report.  Likewise, if the colors do match, then the PID discrepancy identified earlier is due to the Font-Style changing (provided no existing layout \dvs), and thus a Font-Style Violation is reported. Finally, to detect incorrect text content, \GVT utilizes the textual information, preprocessed to remove whitespace and normalize letter cases, and performs a string comparison.  If the strings do not match, then an \textit{Incorrect Text Content} \dv is reported.

\vspace{-0.25cm}
\subsubsection{Detecting Resource Violations}
\label{subsubsec:resource-violations}

	  ~\GVT is able to detect the following resource \dvs: (i) missing component, (ii) extraneous component, (iii) image color, (iv) incorrect images, and (v) component shape. The detection and distinction between \textit{Incorrect Image} \dvs and \textit{Image Color} \dvs requires an analysis that combines two different computer vision techniques. To perform this analysis, cropped images from the mock-up and implementation screenshots according to corresponding \gcs respective bounding boxes are extracted. The goal of this analysis is to determine when the content of image-based \gcs differ, as opposed to only the colors of the \gcs differing. To accomplish this, \GVT leverages PID applied to extracted \gc images converted to a binary color space (\textit{B-PID}) in order to detect differences in \textit{content} and CQ and CH analysis to determine differences in \textit{color} (Sec. \ref{subsubsec:text-violations}). To perform the B-PID procedure, cropped \gc images are converted to a binary color space by extracting pixel intensities, and then applying a binary transformation to the intensity values (\eg converting the images to intensity independent black \& white). Then PID is run on the color-neutral version of these images.  If the images differ by more than an \textit{Image Difference Threshold} ($IDT$), then an \textit{Incorrect Image} \dv (which encompasses the \textit{Component Shape} \dv) is reported. If the component passes the binary PID check, then \GVT utilizes the same CQ and CH processing technique described above to detect \textit{image color} \dvs. Missing and extraneous components are detected as described in Sec. \ref{subsec:gui-comprehension}.

\vspace{-0.25cm}
\subsubsection{Generating Violation Reports}
\label{subsubsec:violation-reports}

	In order to provide developers and designers with effective information regarding the detected \dvs, \GVT generates an \texttt{\small html} report that, for each detected violation contains the following: (i) a natural language description of the design violation(s), (ii) an annotated screenshot of the app implementation, with the affected GUI-component highlighted, (iii) cropped screenshots of the affected \gcs from both the design and implementation screenshots, (iv) links to affected lines of application source code, (v) color information extracted from the CH for \gcs identified to have color mismatches, and (vi) the difference image generated by PID. The source code links are generated by matching the \texttt{\small ids} extracted from the uiautomator information back to their declarations in the layout \texttt{\small xml} files in the source code (\eg those located in the \texttt{\small /res/} directory of an app's source code). We provide examples of generated reports in our online appendix \cite{appendix}. 

\vspace{-0.3cm}
\subsection{Implementation \& Industrial Collaboration}
\label{subsec:impl-collab}

Our implementation of \GVT was developed in Java with a Swing GUI. In addition to running the \GVT analysis the tool executable allows for one-click capture of \texttt{\small uiautomator} files and screenshots from a connected device or emulator.  Several acceptance tests of mock-up/implementation screen pairs with pre-existing violations from apps under development within our industrial partner were used to guide the development of the tool. 12 Periodic releases of binaries for both Windows and Mac were made to deploy the tool to designers and developers within the company.  The authors of this paper held regular bi-weekly meetings with members of the design and development teams to plan features and collect feedback.

	Using the acceptance tests and feedback from our collaborators we tuned the various thresholds and parameters of the tool for best performance. The PID algorithm settings were tuned for sensitivity to capture subtle visual inconsistencies which are then later filtered through additional CV techniques: $F$ was set to $45^{\circ}$, $L$ was set to $100 cdm^2$, and $C$ was set to 1.  The \gc-\textit{Matching Threshold} ($MC$) was set to $1/8th$ the screen width of a target device; the $DT$ for determining suspicious \gcs was set to $20\%$; The $LT$ was set to 5 pixels (based on designer preference); the $CT$ which determines the degree to which colors must match for color-based \dvs was set to $85\%$; and finally, the $IDT$ was set to $20\%$.  \GVT allows for a user to change these settings if desired, additionally users are capable of defining areas of dynamic content (\eg loaded from network activity), which should be ignored by the \GVT analysis.


\vspace{-0.3cm}
\section{Design of the Experiments}
\label{sec:study}

To evaluate \GVTs \textit{performance}, \textit{usefulness} and \textit{applicability}, we perform three complimentary studies answering the following RQs:

\begin{itemize}
	\item \textbf{RQ$_1$}: \textit{How well does \GVT perform in terms of detecting and classifying design violations?} 
	\item \textbf{RQ$_2$}: \textit{What utility can \GVT provide from the viewpoint of Android developers?}
	\item \textbf{RQ$_3$}: \textit{What is the industrial applicability of \GVT in terms of improving the mobile application development workflow?}  
\end{itemize}

	RQ$_1$ and  RQ$_2$ focus on quantitatively measuring the performance of \GVT and the utility it provides to developers through a controlled empirical  and a user study respectively.  RQ$_3$ reports the results of a survey and semi-structured interviews with our collaborators aimed at investigating the industrial  applicability of \GVT.

\vspace{-0.3cm}
\subsection{Study 1: \GVT Effectiveness \& Performance}
\label{subsec:performance-study}

	The \textit{goal} of the first study is to quantitatively measure \GVT in terms of its precision and recall in both detecting and classifying \dvs.

\subsubsection{Study Context}

	To carry out a controlled quantitative study, we manually reverse engineered Sketch mockups for ten screens for eight of the most popular apps on Google Play.  To derive this set, we downloaded the top-10 apps from each category on the Google-Play store removing the various categories corresponding to games (as these have non-standard GUI-components that \GVT does not support).  We then randomly sampled one app from each of the remaining 33 categories, eliminating duplicates (since apps can belong to more than one category).  We then manually collected screenshots and \texttt{\small uiautomator} files from two screens for each application using a Nexus 5, attempting to capture the ``main'' screen that a user would typically interact with, and one secondary screen.  Using the \texttt{\small uiautomator} files, we generated cropped screenshots of all the leaf nodes components for each screen of the app.  From these we were able generate 10 screens from 8 applications that successfully ran through \GVT without any reported violations. 

\vspace{-0.2cm}
\subsubsection{Synthetic \dv Injection}

	With a set of correct mock-ups corresponding to implementation screens in an app, we needed a suitable method to introduce \dvs into our subjects. To this end, we constructed a \textit{synthetic \dv injection tool} that modifies the \texttt{\small uiautomator} \texttt{xml} files and corresponding screenshots in order to introduce design violations from our taxonomy presented in Sec. \ref{sec:motivating-study}. The tool is composed of two components: (i) an \textit{XML Parser} that reads and extracts components from the screen, then (ii) a \textit{Violation Generator} that randomly selects components and injects synthetic violations. We implemented injection for the following types of DVs: 

\noindent \textbf{Location Violation:} The component is moved either horizontally, vertically, or in both directions within the same container. However, the maximum distance from the original point is limited by a quarter of the width of the screen size. This was based on the severity of Layout Violations in our study described in Section \ref{sec:motivating-study}. In order to generate the image we cropped the component and moved it to the new location replacing all the original pixels by the most prominent color from the surroundings in the original location. 

\noindent \textbf{Size Violation:} The component size either increases or decreases by 20\% of the original size. For instances where the component size decreases, we replaced all the pixels by the most prominent color from the surroundings of the original size.

\noindent \textbf{Missing Component Violation:} This violation removes a leaf component from the screen, replacing the original pixels by the most prominent color from the surrounding background. 

\noindent \textbf{Image Violation:} We perturb 40\% of the pixels in an image by randomly generating an RGB value for the pixels affected. 

\noindent \textbf{Image Color Violation:} This rule perturbs the color of an image by shifting the hue of image colors by 30\textdegree. 

\noindent \textbf{Component Color Violation:} This uses the same process as for \textit{Image Color Violations} but we change the color by 180\textdegree. 

\noindent \textbf{Font Violation:} This violation randomly selects a font from the set of: \textsl{Arial}, \textsl{Comic Sans MS}, \textsl{Courier}, \textsl{Roboto}, or \textsl{Times Roman} and applies it to a \texttt{TextView} component. 

\noindent \textbf{Font Color Violation:} changes the text color of a \texttt{TextView} component. We extracted the text color using CH analysis, then we changed the color using same strategy as for \textit{Image Color Violations}.

\vspace{-0.2cm}
\subsubsection{Study Methodology}

	In injecting the synthetic faults, we took several measures to simulate the creation of realistic faults.  First, we delineated 200 different types of design violations according to the distribution defined in our \dv taxonomy in Sec. \ref{sec:motivating-study}.  We then created a pool of 100 screens by creating random copies of the both the \texttt{\small uiautomator} \texttt{\small xml}
files and screenshots from our initial set of 10 screens.  We then used the \textit{synthetic \dv injection tool} to seed faults into the pool of 100 screens according to the following criteria: (i) No screen can contain more than 3 injected \dvs, (ii) each \gc should have a maximum of 1 \dv injected, and (iii) Each screen must have at least 1 injected \dv.  After the \dvs were seeded, each of the 100 screens and 200 \dvs were manually inspected for correctness.  Due to the random nature of the tool, a small number of erroneous \dvs were excluded and regenerated during this process (\eg color perturbed to perceptually similar color.).  The breakdown of injected \dvs is shown in Figure \ref{fig:study1-result}, and the full dataset with description is included in our online appendix \cite{appendix}.

	Once the final set of screens with injected violations was derived, we ran \GVT across these subjects and measured four metrics: (i) detection precision ($DP$), (ii) classification precision ($CP$), (iii) recall ($R$), and (iv) execution time per screen ($ET$).   We make a distinction between detection and classification in our dataset because it is possible that \GVT is capable of detecting, but misclassifying a particular \dv (\eg an \textit{image color \dv} misclassified as an \textit{incorrect image \dv}).  $DP$, $CP$ and $R$ were measured according to the following formulas:
\vspace{-0.1cm}
\begin{equation}
\label{eq:precision}
DP,CP = \frac{T_p}{T_p+F_p}	\quad R = \frac{T_p}{T_p+F_n}
\end{equation}

\noindent where for $DP$, $T_p$ represent injected design violations that were detected, and for $CP$, $T_p$ represents injected violations that were both detected and classified correctly.  In each case $F_p$ correspond to detected \dvs that were either not injected or misclassified.  For Recall, $T_p$ represents injected violations that were correctly detected and $F_n$ represents injected violations that were not detected.  To collect these measures, two authors manually examined the reports from \GVT in order to collect the metrics.

\vspace{-0.2cm}
\subsection{Study 2: \GVT Utility}
\label{subsec:utility-study}

	Since the ultimate goal of an approach like \GVT is to improve the workflow of developers, the \textit{goal} of this second study is to measure the utility (\ie benefit) that \GVT provides to developers by investigating two phenomena: (i) The accuracy and effort of developers in detecting and classifying \dvs, and (ii) the perceived utility of \GVT reports in helping to identify and resolve \dvs. 

\vspace{-0.2cm}
\subsubsection{Study Context}

	We randomly derived two sets of screens to investigate the two phenomena outlined above.  First, we randomly sampled two mutually exclusive sets of 25, and 20 screens respectively from the 100 used in Study 1, ensuring at least one instance of each type of \dv was included in the set.  This resulted in both sets of screens containing 40 design violations in total.  The correct mockup screenshot corresponding to each screen sampled from the study were also extracted, creating pairs of ``correct" mockup and ``incorrect" implementation screenshots. 10 participants with at least 5 years of Android development experience were contacted via email to participate in the survey.

\vspace{-0.2cm}
\subsubsection{Study Methodology}

	We created an online survey with four sections.  In the first section, participants were given background information regarding the definition of \dvs, and the different types of \dvs derived in our taxonomy.  In the second section, participants were asked about demographic information such as programming experience and education level.  In the third section, each participant was exposed to 5 mock-up/ implementation screen pairs (displayed side by side on the survey web page) and asked to identify any observed design violations.  Descriptions of the \dvs were given at the top of this page for reference.  For each screen pair, participants were presented with a dropdown menu to select a type for an observed \dv, and a text field to describe the error in more detail.  For each participant, one of the 5 mock-up screens was a control, containing no injected violations. The 25 screens were assigned to participants such that each screen was observed by two participants and the order of the screens presented to each participant was randomized to avoid bias.  To measure the effectiveness of participants in detecting and describing \dvs, we leverage the $DP$, $CP$ and $R$ metrics introduced in Study 1.  In the fourth section, participants were presented with two screen pairs from the second set of 20 sampled from the user study, as well as the \GVT reports for these screens.  Participants were then asked to answer 5 \textit{user-preferences (UP)} and 5 \textit{user experience (UX)} questions about these reports  which are presented in the following section.  The \textit{UP} questions were developed according to the user experience honeycomb originally developed by Morville \cite{Morville:04} and were posed to participants as free form text entry questions. We forgo a discussion of the free-form question responses due to space limitations, but we offer full anonymized participant responses in our online appendix~\cite{appendix}.  We derived the Likert scale-based \textit{UX} questions using the SUS usability scale by Brooke \cite{Brooke:96}.

\vspace{-0.2cm}
\subsection{Study 3: Industrial Applicability of \GVT}
\label{subsec:performance-study}

\begin{figure}
\centering
\vspace{-0.3cm}
\includegraphics[width=1.1\columnwidth]{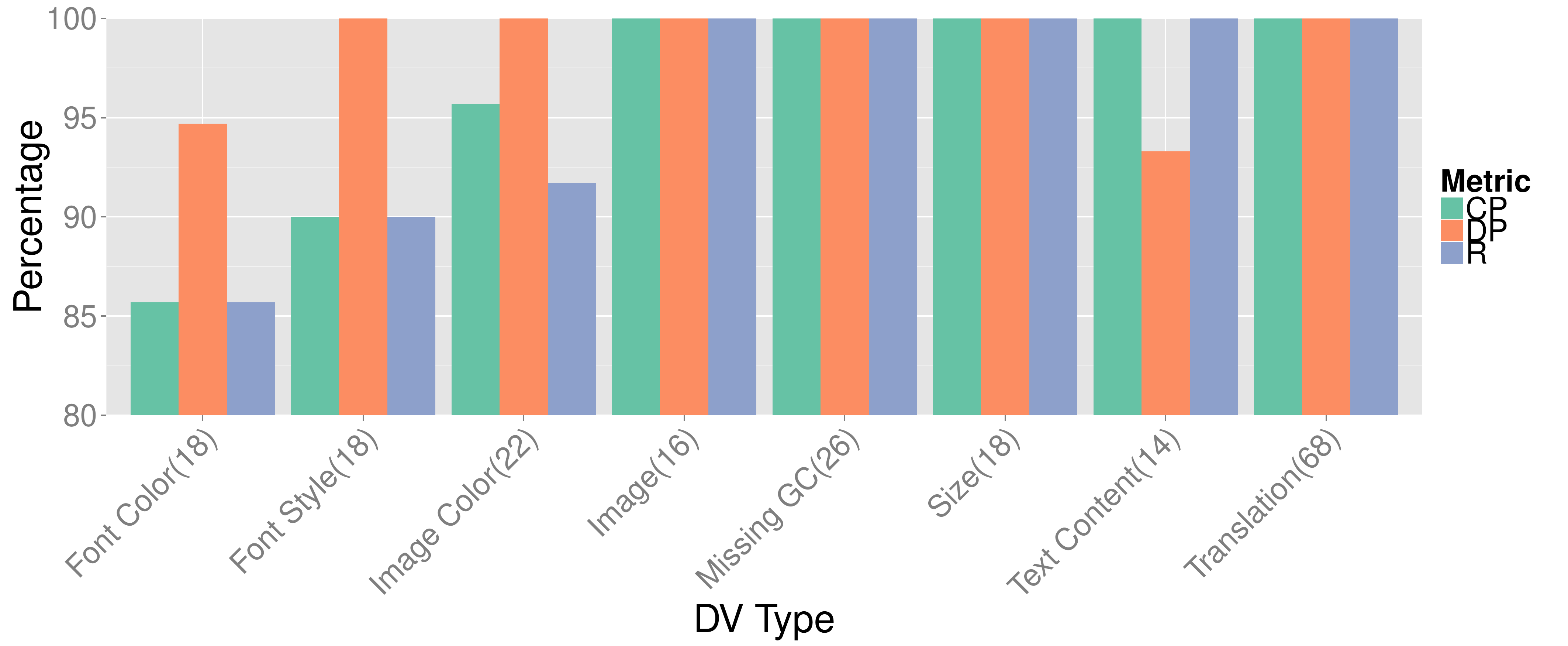}
\vspace{-0.7cm}
\centering
\caption{Study 1 - Detection Precision ($DP$), Classification Precision ($CP$), and Recall ($R$)}
\label{fig:study1-result}
\vspace{-0.1cm}
\end{figure}

	The \textit{goal} of this final study is determine industrial applicability of \GVT.  To investigate this, we worked with Huawei to collect two sources of information: (i) the results of a survey sent to designers and developers who used \GVT in their daily development/design workflow, and (ii) semi-structured interviews with both design and development managers whose teams have adopted the use of \GVT.

\vspace{-0.2cm}
\subsubsection{Study Context \& Methodology}

	We created a survey posing questions related to the \textit{applicability} of \GVT to industrial designers and developers. These questions are shown in Fig. \ref{fig:study3-likert}. The semi-structured interviews were conducted in Chinese, recorded, and then later translated.  During the interview, managers were asked to respond to four questions related to the \textit{impact} and \textit{performance} of the tool in practice. We include discussions of the responses in Section \ref{sec:results} and stipulate full questions in our appendix.


\vspace{-0.2cm}
\section{Empirical Results}
\label{sec:results}

\subsection{Study 1 Results: GVT Performance}
\label{subsec:study1-results}

	The results of Study 1, are shown in Figure \ref{fig:study1-result}. This figure shows the average $DP$, $CP$, and $R$ for each type of seeded violation over the 200 seeded faults and the number of faults seeded into each category (following the distributions of our derived taxonomy) are shown on the x-axis.  Overall, these results are extremely encouraging, with the overall $DP$ achieving $99.4\%$, the average $CP$ being $98.4\%$, and the average $R$ reaching $96.5\%$.  This illustrates that \GVT is capable of detecting seeded faults designed to emulate both the type and distribution of \dvs encountered in industrial settings.  While \GVT achieved at least $85\%$ precision for each type of seeded \dv, it performed worse on some types of violations compared to others.  For instance, \GVT saw its lowest precision values for the \textit{Font-Style} and \textit{Font-Color} violations, typically due to the fact that the magnitude of perturbation for the color or font type was not large enough to surpass the Color or Image Difference Thresholds ($CT$ \& $IDT$). \GVT took 36.8 mins to process and generate reports for the set of 100 screens with injected \dvs, or ~22 sec per screen pair. This execution cost was generally acceptable by our industrial collaborators. 

\vspace{-0.2cm}
\subsection{Study 2 Results: GVT Utility}
\label{subsec:study2- results}

\begin{figure}
\centering
\vspace{-0.3cm}
\includegraphics[width=\columnwidth]{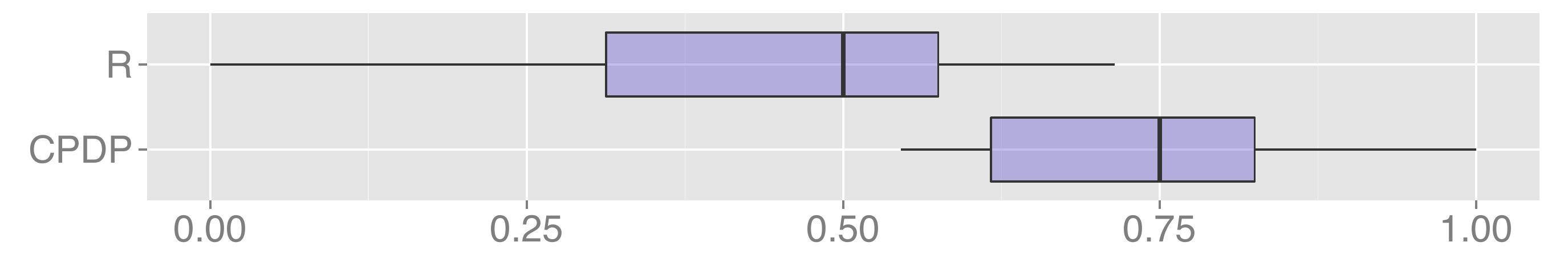}
\vspace{-0.7cm}
\centering
\caption{Study 2 - Developer CP, DP, and R}
\label{fig:study2-dvs}
\vspace{-0.1cm}
\end{figure}

\begin{figure}
\centering
\vspace{-0.4cm}
\includegraphics[width=\columnwidth]{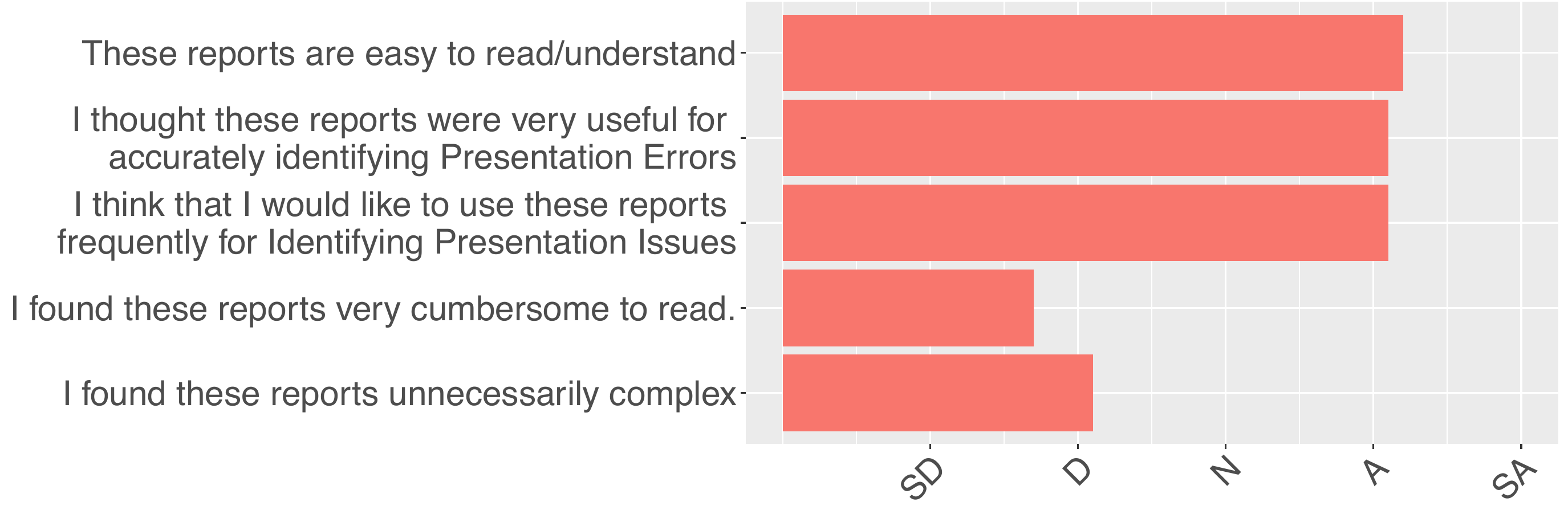}
\vspace{-0.7cm}
\centering
\caption{Study 2 - UX Question Responses. SD=Strongly Disagree, D=Disagree, N=Neutral, A=Agree, SA=Strongly Agree}
\label{fig:study2-likert}
\end{figure}

	The $DP$, $CP$ and $R$ results, representing the Android developers ability to correctly detect and classify \dvs is shown in Figure \ref{fig:study2-dvs} as box-plots across all 10 participants. Here we found $CP$=$DP$, as when a user misclassified violations, they also did not detect them. As this figure shows, the Android developers generally performed much worse compared to \GVT achieving an average $CP$ of under $ \approx 60\%$ and an average $R$ of $\approx 50\%$. The sources of this performance loss for the study participants compared to \GVT was fourfold: (i) participants tended to report minor, acceptable differences in fonts across the examples (despite the instructions clearly stating \textit{not} to report such violations); (ii) users tended to attribute more than one \dv to a single component, specifically for \textit{font style} and \textit{font color} violations despite instructions to report only one; (iii) users tended to misclassify \dvs based on the provided categories (\eg classifying a \textit{layout} \dv for a Text \gc as an \textit{incorrect text} \dv), and (iv) participants missed  reporting many of the injected \dvs, leading to the low recall numbers.  These results indicate that, at the very least, developers can struggle to both detect and classify \dvs between mock-up and implementation screen pairs, signaling the need for an automated system to check for \dvs before implemented apps are sent to a UI/UX team for auditing. This result confirms the notion that developers may not be as sensitive to small \dvs in the GUI as the designers who created the GUI specifications.  Furthermore, this finding is notable, because as part of the iterative process of resolving design violations, designers must communicate to developers \dvs and developers must recognize and understand these \dvs in order to properly resolve them.  This process is often complicated due to ambiguous descriptions of \dvs from designers to developers, or developers disagreeing with designers over the existence or type of a \dv.  In contrast to this fragmented process, \GVT provides clear, unambiguous reports that facilitate communication between designers and developers.

\begin{figure}[t]
\centering
\vspace{-0.3cm}
\includegraphics[width=\columnwidth]{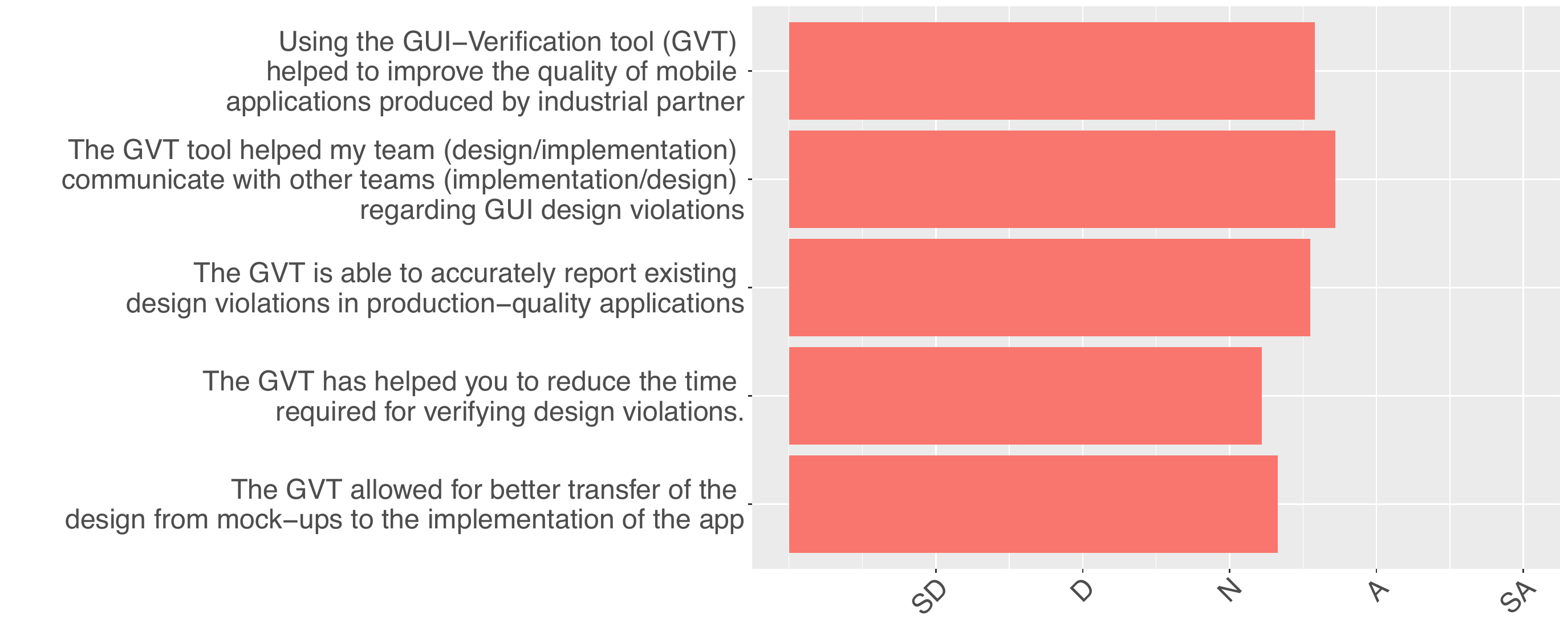}
\vspace{-0.8cm}
\centering
\caption{Study 3 - Applicability Questions. SD=Strongly Disagree, D=Disagree, N=Neutral, A=Agree, SA=Strongly Agree}
\label{fig:study3-likert}
\end{figure}

	Figure \ref{fig:study2-likert} illustrates the responses to the likert based UX questions, and the results are quite encouraging.  In general, participants found that the reports from \GVT were easy to read, useful for identifying \dvs and indicated that they would like to use the reports for identifying \dvs. Participants also indicated that the reports were not unnecessarily complex or difficult to read. We asked the participants about their preferences for the \GVT reports as well, asking about the most and least useful information in the reports.  \textit{Every single} participant indicated that the highlighted annotations on the screenshots in the report were the most useful element. Whereas most users tended to dislike the PID output included at the bottom of the report, citing this information as difficult to comprehend.

\vspace{-0.3cm}
\subsection{Study 3 Results: Industrial Applicability}
\label{subsec:study3-results}

	The results for the \textit{applicability} questions asked to 20 designers and developers who use \GVT in their daily activities is shown in Figure \ref{fig:study3-likert}. A positive outcome for each of these statements correlates to responses indicating that developers ``agree'' or ``strongly agree''.  The results of this study indicate a weak agreement of developers for these statements, indicating that while \GVT is generally applicable, there are some drawbacks that prevented developers and designers from giving the tool unequivocal support. We explore these drawbacks by conducting semi-structured interviews.

	In conducting the interviews, one of the authors asked the questions presented in Figure \ref{fig:study3-likert} to 3 managers (2 from UI/UX teams and 1 from a Front-End development team). When asked whether \GVT contributed to an increased quality of mobile applications at the company, all three managers tended to agree that this was the case. For instance, one of the design managers stated, \textit{``Certainly yes. The tool is the industry's first"} and the other designer manager added, \textit{``When the page is more complicated, the tool is more helpful"}.

	When asked about the overall performance and accuracy of the tool in detecting \dvs, the manager from the implementation team admitted that the current detection performance of the tool is good, but suggested that dynamic detection of some components may improve it, stating, \textit{``[\dvs] can be detected pretty well... [but the tool is] not very flexible. For example, a switch component in the design is open, but the switch is off in the implementation"}. He suggested that properly handling cases such as this would make the tool more useful from a developers perspective. One of the design team managers held a similar view stating that, \textit{``Currently, most errors are layout errors, so tool is accurate. Static components are basically detected, [but] maybe the next extension should focus on dynamic components."}  While the current version of the \GVT allows for the exclusion of regions with dynamic components, it is clear that both design and development teams would appreciate proper detection of \dvs for dynamic components. Additionally, two of the managers commented on the ``\textit{rigidity}'' of the \GVTs current interface, and explained that a more streamlined UI would help improve its utility.

	When asked about whether \GVT improved communication between the design and development teams, the development team manager felt that while the tool has not improved communication yet, it did have the potential to do so, \textit{``At present there is no [improvement] but certainly there is the potential possibility."} The design managers generally stated that the tool has helped with communication, particularly in clarifying subtle \dvs that may have caused arguments between teams in the past, \textit{``If you consider the time savings on discussion and arguments between the two teams, this tool saves us a lot of time"}. Another designer indicated that the tool is helpful at describing \dvs to developers who may not be able to recognize them with the naked eye \textit{``We found that the tool can indeed detect something that the naked eye cannot"}.  While there are certainly further refinements that can be made to \GVT, it is clear that the tool has begun to have a positive impact of the development of mobile apps, and as the tool evolves within the company, should allow for continued improvements in quality and time saved.


\vspace{-0.3cm}
\section{Limitations \& Threats to Validity}
\label{sec:limitations}

\textit{\textbf{Limitations}}: While we have illustrated that \GVT is applicable in an industrial setting, the tool is not without its limitations.  Currently, the tool imposes lightweight restrictions on designers creating Sketch mock-ups, chief among these being the requirement that bounding boxes of components do not overlap.  Currently, \GVT will try to resolve such cases during the \textit{GUI-Comprehension stage} using an Intersection over union (IOU) metric.  

\noindent \textit{\textbf{Internal Validity}}: While deriving the taxonomy of \dvs, mistakes in classification arising from subjectiveness may have introduced unexpected coding. To mitigate this threat we followed a set methodology, merged coding results, and performed conflict resolution.

\noindent \textit{\textbf{Construct Validity}}: In our initial study (Sec. \ref{sec:motivating-study}), a threat to construct validity arises in the form of the manner in which coders were exposed to presentation failures.  To mitigate this threat, designers from our industrial partner manually annotated the screen pairs in order to clearly illustrate the affected \gcs on the screen. In our evaluation of \GVT threats arise from our method of \dv injection using the \textit{synthetic fault injection tool}.  However, we designed this tool to inject faults based upon both the type and distribution of faults from our \dv taxonomy to mitigate this threat.  

\noindent \textit{\textbf{External Validity}}: In our initial study related to the \dv taxonomy, we utilized a dataset from a single (albeit large) company with examples across several different applications and screens. There is the potential that this may not generalize to other industrial mobile application development environments and platforms or mobile app development in general. However given the relatively consistent design paradigms of mobile apps, we expect the categories and the sub-categories within the taxonomy to hold, although it is possible that the distribution across these categories may vary across application development for different domains. In Study 3 we surveyed employees at a single (though large) company, and findings may differ in similar studies at other companies.


\vspace{-0.3cm}
\section{Related Work}
\label{sec:related-work}
\vspace{-0.1cm}

\noindent \textit{\textbf{Web Presentation Failures}}: The work most closely related to our approach are approaches that aim at detecting, classifying and fixing presentation failures in web applications \cite{Mahajan:ICST15,Mahajan:ICST16,RoyChoudhary:ICSE13,Mahajan:ISSTA17}. In comparison to these approaches, \GVT also performs detection and localization of presentation failures, but is the first to do so for mobile apps. In addition to the engineering challenges associated with building an approach to detect presentation failures in the mobile domain (\eg collection and processing of GUI-related data) \GVT is the first approach to leverage metadata from software mock-up artifacts (\eg Marketch) to perform \gc matching based upon the spatial information collected from both mock-ups and dynamic application screens, allowing for precise detection of the different types of \dvs delineated in our industrial \dv taxonomy.  \GVT is also the first to apply the processes of CQ, CH analysis, and B-PID toward detecting differences in the content and color of icons and images displayed in mobile apps.  \GVT also explicitly identifies and reports different faulty properties (such as errors in component location, or text).

\noindent \textit{\textbf{Cross Browser Testing}}: Approaches for XBT (or cross browser testing) by Roy Choudhry \textsl{et. al.} \cite{RoyChoudhary:ICSE13,Choudhary:ICST12,RoyChoudhary:ICSM10} examine and automatically report differences in web pages rendered in multiple browsers. These approaches are currently not directly applicable to mock-up driven development for mobile apps. 

	\noindent \textit{\textbf{Visual GUI Testing}}: A concept known as Visual GUI Testing (VGT) aims to test certain visual aspects of a software application's GUI as well as the underlying functional properties.  To accomplish this visual GUI testing usually executes actions on a target applications in order to exercise app functionality \cite{Alegroth:ICST15,Nguyen:ASE14,Alegroth:EMSE17,ISSA:WSE12}. In contrast to these approaches, \GVT is designed to apply to mobile-specific \dvs, is tailored for the mock-up driven development practice, and is aimed \textit{only} at verifying visual properties of a mobile app's GUI. 

	\noindent \textit{\textbf{Other Approaches}}: There are other approaches and techniques that related to identifying problems or differences with GUIs of mobile apps.  Xie \textsl{et al.} introduced GUIDE \cite{Xie:ICSM09}, a tool for GUI differencing between successive releases of GUIs for an app by matching components between GUI-hierarchies.  \GVT utilizes a matching procedure for leaf node components as direct tree comparisons are not possible in the context of mock-up driven development. There has also been both commercial and academic work related to graphical software built specifically for creating high-fidelity mobile app mock-ups or mockups that encode information for automated creation of code for a target platform \cite{Meskens:EICS09, mockup-io,fluid-ui,proto-io}.  However, such tools tend to either impose too many restrictions on designers or do not allow for direct creation of code, thus \dvs still persist in practice.

\vspace{-0.4cm}
\section{Conclusion \& Future Work}
\label{sec:conclusion}
\vspace{-0.1cm}

	In this paper, we have formalized the problem of detecting design violations in mobile apps, and derived a taxonomy of design violations based on a robust industrial dataset.  We presented \GVT, an approach for automatically detecting, classifying, and reporting design violations in mobile apps, and conducted a wide ranging study that measured performance, utility, and industrial applicability of this tool.  Our results indicate that \GVT is effective in practice, offers utility for developers, and is applicable in industrial contexts.

\vspace{-0.3cm}
\begin{acks}
\vspace{-0.1cm}
The authors would like to thank Kebing Xie, Roozbeh Farahbod, and the developers and designers at Huawei for their support.
\end{acks}

\balance
\bibliography{ms}
\bibliographystyle{abbrv}

\end{document}